\documentclass[]{article}

\usepackage[a4paper, left=0.6in, right=0.6in, top=0.6in, bottom=0.6in, footskip=.25in]{geometry}
\usepackage{authblk}
\usepackage{graphicx}
\usepackage{multirow}

\usepackage{natbib}

\title{Micro-scale Graded Mechanical Metamaterials Exhibiting Versatile Poisson's Ratio}

\author[1,2]{K. K. Dudek*}
\author[3]{L. Mizzi}
\author[1]{J. A. Iglesias Mart\'{\i}nez}
\author[3]{A. Spaggiari}
\author[1]{G. Ulliac}
\author[4,5]{R. Gatt}
\author[4,6]{J. N. Grima}
\author[1]{V. Laude}
\author[1]{M. Kadic*}

\affil[1]{Institut FEMTO-ST, CNRS, Université Bourgogne Franche-Comté, Besançon 25030, France} 
\affil[2]{Institute of Physics, University of Zielona Gora, ul. Szafrana 4a, Zielona Gora 65-069, Poland}
\affil[3]{Department of Engineering Sciences and Methods, University of Modena and Reggio Emilia, Reggio Emilia, Italy}
\affil[4]{Metamaterials Unit, Faculty of Science, University of Malta, Msida MSD 2080, Malta}
\affil[5]{Centre for Molecular Medicine and Biobanking, University of Malta, Msida, MSD 2080, Malta}
\affil[6]{Department of Chemistry, Faculty of Science, University of Malta, Msida MSD 2080, Malta}

\date{}

\begin{document}
\maketitle

\begin{abstract}
	The ability to control Poisson's ratio of functional materials has been one of the main objectives of researchers attempting to develop structures efficient from the perspective of protective, biomedical and soundproofing devices. This task becomes even more challenging at small scales, such as the microscale, where the possibility to control mechanical properties of functional materials is very significant, like in the case of flexible electronics. In this work, we propose novel microscopic 2D and 3D functionally-graded mechanical metamaterials capable of exhibiting a broad range of Poisson's ratio depending on their composition. More specifically, we show that upon adjusting the number of structural elements corresponding to one type of the substructure at the expense of another, it is possible to change the resultant Poisson's ratio of the entire system from highly positive to highly negative values as well as to achieve arbitrary intermediate values. Finally, in addition to static properties, we also analyze dynamic properties of the considered structures. Namely, we show that the variation in the composition of considered mechanical metamaterials affects the velocity of a wave propagating through the system. This, in turn, could be essential in the case of applications utilizing localized wave attenuation or sensors.
\end{abstract}

\section{Introduction}

In recent years, functional materials have been the subject of intense studies due to their ability to exhibit versatile mechanical as well as other types of behavior. Particularly important in this regard is the specific class of functional materials called mechanical metamaterials \cite{Florijn2014, Neville_2016, Coulais_Teomy_2016, Coulais2018, Wenz_2021, christensen2015vibrant, kadic20193d, Chen_Kadic_2020, Frenzel_Wegener_2017, Jiang_Chen_2020, Yao_Su_2022, Wang_Luo_2014, Gao_Xue_2019, Placidi_Rosi_2013, Wang_Zhao_2020, Zheng_Uto_2022, Wu_Wang_2020, doi:10.1126/science.1260960, doi:10.1126/sciadv.aat8313, doi:10.1177/07316844211009599, HAMZEHEI2020103291, SUN2020114615} which stems from their continuously growing applicability in numerous industries. In general, mechanical metamaterials are rationally designed structures that can exhibit unusual mechanical properties such as auxetic behavior (negative Poisson's ratio) \cite{Wojciechowski_1989, Lakes1987, Evans_Alderson_2000, Babaee_2013, Novak_Biasetto_2021, Lim_2019_compos, Jiang_Ren_2022, Grima_squares_2000, Grima_triangles3, Albertini2021, Novak_Duncan_2021, Ruzzene_Mazzarella_2022, Hou_Neville_2014}, negative stiffness \cite{Hewage2016, Tan_Wang2022, Chen_Tan_2021, Bertoldi_Vitelli_2017} as well as negative compressibility \cite{Nicolau_Motter_2012, Baughman1998, Dudek_triangles_2016, Lakes_Wojciechowski_2008}. As demonstrated over the last thirty years, these counterintuitive mechanical properties can be utilized in a plethora of applications ranging from protective \cite{Miniaci2016, Imbalzano2018} to biomedical \cite{Kolken_Zadpoor2018, Teunis_van_Manen_2021, Wang_Wu_2016, Xue_Saha_2021} and soundproofing devices \cite{Quadrelli2021, Guenneau_Ramakrishna_2007, Fleury2014, Fleury2016}. Recently, it has been also possible to observe a rapid increase in the commercial appeal of mechanical metamaterials capable of exhibiting such properties in the case of easily accessible products that are not necessarily devoted to high-technology industries. A prime example of such an approach are auxetic mechanical metamaterials used in the design of sports equipment \cite{Duncan_sport_2018}. 

Despite numerous advantages of standard mechanical metamaterials, it is necessary to note that they often share several limitations. Mechanical metamaterials are typically designed in a way where the entire structure is composed of one type of geometry corresponding to a specific mechanical characteristic \cite{Grima_squares_2000}. As a result, mechanical properties of such structures cannot normally be significantly modified which in some cases reduces the applicability of the system. In order to overcome this limitation, a number of new approaches were proposed that allow to significantly change mechanical properties of the system even after it is manufactured \cite{Rafsanjani_2018}. A prime example of such approach corresponds to the use of active composite metamaterials that can have their behavior controlled via the magnitude of the external stimulus for example in the form of the change in the temperature or the electric and magnetic field \cite{Galea_Dudek_2021, Korpas_2021, Xin_Liu_2020, Dudek_Julio_magnetic_Adv_Mater_2023}. Another very interesting example corresponds to hierarchical mechanical metamaterials \cite{Lakes_hierarchical_1993, Gatt_hierarchical_2015, Cho_hierarchical_2014, Tang_Lin_hierarchical_2015, An_Bertoldi_2020, Oftadeh_2014, Dudek_2022_Adv_Mater, Zhang_Scarpa_2021, Dudek_pssb_2022}, i.e. systems composed of substructures that can deform irrespective from the rest of the structure. As reported in recent years, hierarchical mechanical metamaterials can exhibit very versatile mechanical behavior that can be controlled throughout the deformation process \cite{An_Bertoldi_2020, Dudek_2022_Adv_Mater}. Nevertheless, despite their undeniable appeal, the manufacturing process of such active mechanical metamaterials is often challenging and expensive relative to the standard mechanical metamaterials. In addition, to fully showcase their tunable properties, the aforementioned active mechanical metamaterials often require access to equipment that is not accessible outside research facilities \cite{Galea_Dudek_2021}. Hence, it is relatively difficult to make applications related to these structures to be easily available to the general public. Nevertheless, it is also possible to observe a broad range of the mechanical behavior in the case of standard mechanical metamaterials \cite{Mizzi_Mahdi_2018}. One of the possible ways of achieving it corresponds to the use of several different substructures in the design of the system \cite{Liu_Tao_2020, Guo_2021, Kolken_Janbaz_2018, Mirzaali_Zadpoor_2020, Dudek_magnetic_bumper_2019}.   

In the literature, there are relatively few examples of graded mechanical metamaterials composed of several substructures \cite{Mirzaali_Zadpoor_2020, Guo_Huang_2021, Han_Kei_2022, NIKNAM2020109129, SHAO2021114366, PAGLIOCCA2022110446, YAO2021113313}. However, despite this direction of studies being still at its infancy, there are already several interesting papers related to this concept. More specifically, in several relatively recent studies \cite{Liu_Tao_2020, Guo_2021, Kolken_Janbaz_2018, BOLDRIN2016114}, it has been shown that the resultant mechanical property of a graded structure (e.g. Poisson's ratio) can be different from the properties of its constituent substructures and typically depends on the interplay between them. In addition, it has also been demonstrated that multi-substructure mechanical metamaterials can exhibit shape morphing \cite{Mirzaali_Zadpoor_2018}. Nevertheless, some of the studies related to this concept correspond to graded metamaterials where the resultant structures are relatively unstable and can lose their ability to exhibit the desired mechanical properties if one deforms the structure in an arbitrary direction (especially in the case of graded composites). This in turn significantly reduces their applicability. In addition, it should be noted that most of the reported multi-substructure systems are only two-dimensional which does not allow to replicate their behavior in 3D. Finally, it should be emphasized that multi-substructure mechanical metamaterials known in the literature are typically constructed at the macroscale. On the other hand, in the case of numerous applications including amongst others flexible electronics and biomedical devices, it is essential to use mechanical metamaterials capable of exhibiting a broad range of mechanical properties at much lower scales including the micro-scale.

In this work, we propose a novel 2D and 3D functionally-graded mechanical metamaterial composed of multiple substructures that can exhibit a broad range of the Poisson's ratio. Due to the specific selection of substructures as well as their connectivity, the deformation of the analyzed system depends primarily on one parameter. This leads to a high stability of the structure and allows to ensure that its deformation will not be significantly affected by the choice of the direction in which it is being deformed. We also show that the considered system can be constructed at the micro-scale in order to prove its suitability for applications where the ability to exhibit specific mechanical properties at small scales is essential.

\section{Analyzed structures}
\begin{figure*}
	\centering
	\includegraphics[width=0.9\textwidth]{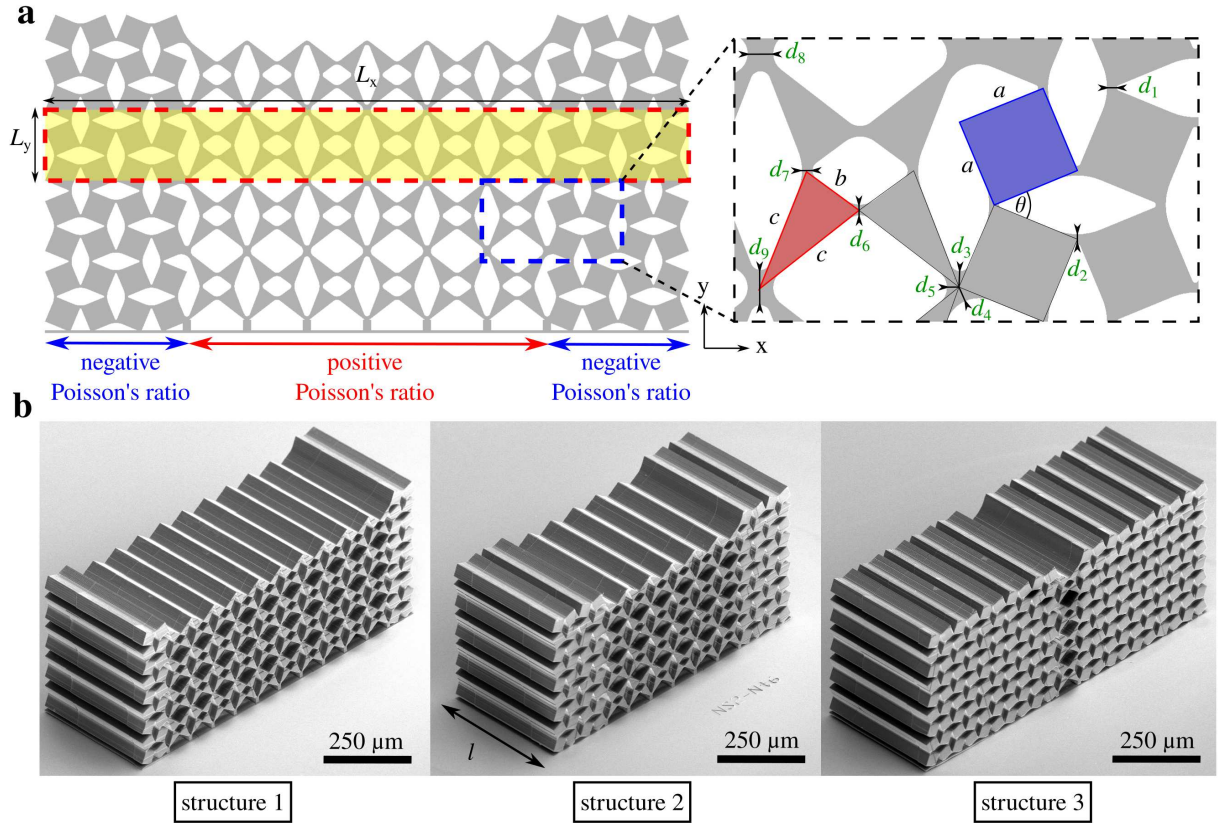}
	\caption{{\bf a)} The graphical representation of the cross-section of an example of the considered system. The highlighted part of the structure corresponds to the unit-cell of the system. {\bf b)} SEM (Scanning Electron Microscopy) images of experimental prototypes of the three examples (structure 1: $N_{\rm{S,x}}$ = 1, $N_{\rm{T,x}}$ = 10, structure 2: $N_{\rm{S,x}}$ = 2, $N_{\rm{T,x}}$ = 6, structure 3: $N_{\rm{S,x}}$ = 5, $N_{\rm{T,x}}$ = 1) of quasi-2D structures analyzed in this work. The overall $x$-dimension for the three considered structures is equal to 1.09 mm, 0.95 mm and 1.13 mm respectively. For all structures, their $y$ and $z$ dimensions are the same and are equal to 0.47 mm and 0.4 mm.}
	\label{model_fig}
\end{figure*}

In this work, we propose a novel mechanical metamaterial composed of two different substructures corresponding to very different values of the Poisson's ratio. Thanks to the implementation of multiple substructures associated with a very different mechanical profile, we aim to show that depending on the ratio of a number of structural elements corresponding to each of the substructures, it is possible to significantly adjust the resultant properties of the graded metamaterial. The two mechanical metamaterials selected in this work as substructures correspond to the rotating squares system \cite{Grima_squares_2000} and the rotating triangles system with four triangles forming its rectangular unit-cell \cite{Milton_expanders_2013, Dudek_triangles_2016} (in the literature there are also other types of mechanical metamaterials based on rotating triangles \cite{Grima_triangles3}). The first of these 2D substructures is known to exhibit auxetic behavior corresponding to the Poisson's ratio equal to -1 irrespective of the loading direction in its idealized form. On the other hand, as shown in the literature \cite{Dudek_triangles_2016}, the later system based on isosceles triangle motifs always exhibits a positive Poisson's ratio along the principal axes. This means that in general, one should expect that the mechanical metamaterial composed primarily of the rotating square-based elements should exhibit the negative Poisson's ratio while the graded metamaterial consisting mainly of the rotating triangles will exhibit much larger values of the Poisson's that may result in the positive values of this parameter.

As shown in Fig. \ref{model_fig}(a), the quasi-2D version of the considered mechanical metamaterial is defined in a way where two identical auxetic substructures are separated by the positive Poisson's ratio substructure corresponding to the rotating triangles. In order to quantitatively describe the amount of each substructure within the system, we introduce variables $N_{\rm{S,x}}$ and $N_{\rm{T,x}}$, where $N_{\rm{S,x}}$ stands for the number of square-based unit-cells ($2 \times 2$ squares) on one side of the system (measured along the $x$-axis). Similarly, $N_{\rm{T,x}}$ corresponds to the number of triangle-based unit-cells (2 $\times$ 2 triangles) measured along the $x$-direction. Thus, in the case of the example presented in Fig. \ref{model_fig}(a), one can write down the following: $N_{\rm{S,x}}$ = 2 and $N_{\rm{T,x}}$ = 6. 


In terms of the geometric parameters, for all types of structures investigated in this study, the rotating squares used in the auxetic substructure have the side length of $a$ = 40 $\mu$m while the isosceles triangles have dimensions $b$ = 30 $\mu$m and $c$ = 56 $\mu$m (see Fig. \ref{model_fig}(a)). As shown in Fig. \ref{model_fig}(a), the structural elements are connected to each other at vertices by means of hinges having the non-zero thickness that varies depending on the position within the graded metamaterial. It is also important to note that since both types of substructures are `unimode' systems, i.e. their deformation can be described solely by means of one parameter corresponding to the angle between structural elements, the resultant structures analyzed in this study are also `unimode' systems. Thus, the deformation of the entire graded structure like the one shown in Fig. \ref{model_fig}(a) can be described through the change in only one parameter. In this study, this parameter was selected as the angle between the adjacent rotating square elements and is denoted as $\theta$. Using this variable, one can derive analytical expressions corresponding to the linear dimensions of the unit-cell of the entire system. This unit-cell is highlighted by means of the yellow color in Fig. \ref{model_fig}(a). Furthermore, upon assuming an idealized scenario where respective squares and triangles are connected to each other at vertices in a pin-jointed manner (i.e. freely rotating structural elements with no flexure of the material), the linear dimensions of such unit-cell can be defined as follows:

\begin{equation}
L_{x} = 4 N_{\rm{S,x}} a \left[\sin \left(\frac{\theta}{2} \right) + \cos \left(\frac{\theta}{2} \right) \right] + 2 N_{\rm{T,x}} \left[c \cos \left(\frac{\gamma}{2} \right) + b \cos \left(\frac{\theta_{1}}{2} \right) \right]
\end{equation}

\begin{equation}
L_{y} = 2 a \left[\sin \left(\frac{\theta}{2} \right) + \cos \left(\frac{\theta}{2} \right)\right]
\end{equation}  

\noindent
where, $\gamma = \arccos \left(\left(2c^{2} - L_{y}^{2} \right) / \left(2c^{2} \right) \right)$, $\theta_{1}$ = $\pi$ - $\gamma$ + $\beta$ and $\beta$ = $\arccos$ $\left(\left(2c^{2} - b^{2} \right) / \left(2c^{2} \right) \right)$. In addition, it is possible to determine the maximum range of the $\theta$ angle corresponding to the analyzed geometry (i.e. the range of angles associated with configurations that can be assumed by the system without the structural elements overlapping):

\begin{equation}
\theta_{min} = 0^{\circ} \, \, \mathrm{or} \, \, \theta_{min} = \arcsin \left[\left(\frac{2}{a c} \right)^{2} s (s - c)^{2} (s - b) - 1  \right] 
\end{equation}

\begin{equation}
\theta_{max} = \arcsin \left( \left(\frac{c}{a} \right)^{2} - 1 \right) \, \, \rm{or} \, \, \theta_{max} = 90^{\circ} 
\end{equation}

\noindent
where, $s = (2c + b) / 2$. At this point, it is important to note that the above analytical model is only applicable if geometric parameters allow the system to assume the geometry with structural elements connected as shown in Fig. \ref{model_fig}(a). It is also important to mention that all of the quasi-2D structures considered in this work have a non-zero thickness equal to 400 $\mu$m that is denoted by means of the $l$ variable in Fig. \ref{model_fig}(b). In addition, it should be emphasized that, in the latter part of this work, we also investigate the behavior of three-dimensional structures inspired by quasi-2D structures presented in Fig. \ref{model_fig}(b).


\section{Materials and Methods}
\subsection{Fabrication}
In order to produce experimental micro-scale prototypes investigated in this work, the commercial 3D printer (Photonic Professional GT+, Nanoscribe GmbH) operating based on the two-photon lithography method was used. The negative tone IP-S photoresin (Nanoscribe GmbH) was used to manufacture the samples that enabled us to print structural elements with connections not smaller than 2 $\mu$m. A drop of IP-S resin was deposited on an ITO-coated soda-lime glass substrate (dimensions: 25 × 25 × 0.7 mm$^{3}$) and then photopolymerized through a 25X-objective with a femtosecond laser operating at $\lambda$ = 780 nm. The standard printing parameters are slicing and hatching distances equal to 1 $\mu$m and 0.5 $\mu$m respectively, a laser power of 90\% and a galvanometric scanning speed of 100 mm/s. During the printing process, respective unit-cells were polymerized individually while ensuring that there is a small overlap between them. After the printing process, the sample was developed for the duration of 25 min in the PGMEA solution (Propylene Glycol Methyl Ether acetate). This procedure was conducted to remove the unexposed photoresist. Finally, the sample was rinsed for 3 min in isopropyl alcohol (IPA) in order to clean it from the aforementioned developer solution. At this point, it is important to note that changing values of the above parameters such as the laser power may significantly affect the quality of the printed samples. This is particularly important in terms of the hinges connecting the adjacent structural elements where the change in one of these parameters may significantly affect their thickness. Hence, the performance of the structure could be modified relative to computer simulations where the desired thickness of hinges would be used.

\subsection{Experiment}
In this work, to test the mechanical properties of micro-scale prototypes corresponding to the considered graded laminate-like mechanical metamaterials, the samples were compressed along the $y$-direction (see Fig. \ref{results_2D}) at the constant speed of 1 $\mu$m s$^{-1}$. To this aim, an external indenter in the form of a polished screw was used that had a circular cross-section having a diameter of 3 mm. It is important to note that this dimension is significantly larger than the longest dimension of the quasi-2D structures (1.09 mm, 0.95 mm, and 1.13 mm along the x-axis for structure 1, structure 2, and structure 3 respectively). Hence, all elements constituting the top part of a given system were compressed simultaneously during the deformation. In addition to quasi-2D samples, as explained in the further part of this work, a 3D version of the considered structures was also investigated. In this case, 3D samples were subjected to the same compressive conditions as was the case for quasi-2D prototypes. Namely, through the use of a large indenter (3 mm in diameter) that covers the entire surface of a sample, 3D structures were compressed along the y-axis at the constant rate of 1 $\mu$m s$^{-1}$. At this point, it is important to emphasize that throughout the compression process of both quasi-2D and 3D prototypes, the bottom part of all of the analyzed samples was constrained from any movement since samples were not removed from the substrate after the printing process.

To analyze Poisson's ratio of finite quasi-2D structures presented in Fig. \ref{model_fig}(b), only a specific part of the sample was taken into consideration, i.e. the highlighted unit-cell in Fig. \ref{model_fig}(a). For these structures, Poisson's ratio was calculated based on the average outline of the considered cell throughout its deformation process. More specifically, positions of the outmost vertices constituting the initial undeformed unit-cell were tracked throughout the deformation process to determine the average vertical and horizontal bounds of the rectangular region associated with the unit-cell. The variation in the shape of such rectangular region was later used to determine the strains as well as the corresponding Poisson's ratio. A detailed explanation of the selection of the specific points within the unit-cell used for such analysis is provided in the Supplementary Information. It should be also emphasized that the selection of the specific unit-cell presented in Fig. \ref{model_fig}(b) was not coincidental. This stems from the fact that this specific unit-cell was relatively far away from the constrained bottom of the system so that its behavior was not be significantly affected by it. At the same time, one could have not selected the topmost unit-cell to conduct the quantitative analysis since such unit-cell was expected to exhibit significant edge effects and would have been in direct contact with the indenter. Of course, this approach does not resolve the issue of free edges of the considered unit-cell on the left and right hand side of the investigated region. To avoid it, it would be normally necessary to consider the structure composed of several unit-cells in each row of the structure. However, given the size and complexity of the analysed unit-cell, this would be extremely expensive and very difficult to achieve during the manufacturing process at the micro-scale. Because of this, in order to provide information about mechanical properties of the system unaffected by the edge effects, in the later part of this work, we also provide results corresponding to unit-cells with periodic boundary conditions. 

In the case of 3D structures, Poisson's ratio was calculated based on the outline of the entire structure rather than the specific unit-cell. This stems from the fact that the considered 3D structures have a much more complex shape than their quasi-2D counterparts which makes it very difficult to track the deformation of a specific part of the system. More specifically, since the analyzed samples were fabricated at the micro-scale, cameras must operate at a very small working distance and can capture a focused image of only those elements that are located approximately at the same distance away from its lens. In the case of considered 3D structures, specific parts of a sample are located at very different distances from the lens. Thus, the best approach to estimate Poisson's ratio is to focus the camera image on the outmost edges of the entire system while neglecting the behavior within the structure. As a result, the edges of 3D samples at the micro-scale can be easily distinguished from the background. It should be also noted that to record the deformation process, the optical camera corresponding to the 10$\times$ (3D prototypes) and 20$\times$ (quasi-2D structures) magnification factor was used.

\subsection{FEM simulations}
In this work, to verify experimental results and conduct other types of simulations, the Finite Element Method (FEM) was used. First of all, the behavior of the quasi-2D models matching exactly experimental prototypes (see Fig. \ref{model_fig}(b)) were simulated through the use of the commercial software COMSOL Multiphysics. To this aim, the boundary conditions were defined in a way so that the resultant behavior of the model would be as similar to the experiment as possible. More specifically, the bottommost parts of the considered models were fixed in space while the other elements of the system remained unconstrained. Then, to induce the deformation process, a rigid, undeformable plate was used to compress the structure. Contact elements were applied to the surface edge nodes of the system as well as to internal surfaces, where necessary, in order to replicate as accurately as possible the experimental compression test. Furthermore, in the conducted simulations, the use of geometric nonlinearities was assumed. It was also assumed that the material corresponding to the investigated samples was isotropic and its linear mechanical properties, in the form of the Young's modulus and Poisson's ratio, were defined as 4 GPa and 0.4 respectively, i.e. mechanical properties matching the polymerized resin used to produce the experimental prototypes. It should be also mentioned that the element type used in the conducted simulations was CPS4R four-node plane stress element. In addition, the average dimension of an element was equal to 3 $\mu$m with the total number of elements per unit-cell being equal to 34875, 60396, and 71676 respectively for the three considered types of structures. The solver used in the conducted simulations utilized the Quasi-Newton Method (static Implicit analysis). It should be also noted that in the case of the anlysis of the dynamic properties of the system, approximately the same average size of elements was used. Most importantly, the size of elements at the edge of the unit-cell had a size of around 0.3 $\mu$m.

In addition to FEM simulations related to finite systems, the behavior of individual unit-cells was also investigated through FEM computer simulations to compare it with the analytical model derived in this work and gain a better understanding of how cells within much larger systems than those considered in this study could behave. To this aim, the same material properties as those defined for the finite models were implemented. However, in this case, instead of constraining the bottom part of the system, periodic boundary conditions were implemented following the method described in \cite{Mizzi_PBCs} (see Supplementary Information). Furthermore, to obtain realistic results, the geometric nonlinearity was used in our simulations. Finally, it is worth mentioning that for all of the FEM simulations conducted in this study, the convergence test was conducted to determine the appropriate mesh size.   

\subsection{Parameters}
In addition to the already defined parameters, it is important to mention that the thicknesses of hinges defined in Fig. \ref{model_fig}(a) corresponded to initial values shown in Table \ref{initial_parameters}. For all structures considered in this work, the initial value of the angle $\theta$ was set to be equal to 45$^{\circ}$. 

\begin{table}
	\renewcommand{\arraystretch}{1.35}
	\begin{tabular}{ |c|c|c|c|c|c|c|c|c|c|}	
		
		\hline
		& $d_{\rm{1}}$ & $d_{\rm{2}}$ & $d_{\rm{3}}$ & $d_{\rm{4}}$ & $d_{\rm{5}}$ & $d_{\rm{6}}$ & $d_{\rm{7}}$ & $d_{\rm{8}}$ & $d_{\rm{9}}$\\ \hline
		$\rm{S_{1}}$ & 4.2     &  4.2    & 6.0     &  6.0    &  4.0    &  3.2    &  3.8    &  8.0    & 12.0   \\ \hline
		$\rm{S_{2}}$ &  4.4    &  4.4    &   7.4   &   6.5   &   5.7   &  4.8    &  4.7    &  11.4    & 15.8   \\ \hline
		$\rm{S_{3}}$ &  3.6    &  3.6    &   7.5   &  7.0    &  5.5    &  6.8    &   6.6   &  11.0    &  12.0  \\ \hline
		
	\end{tabular}
	\caption{Thickness of hinges expressed in $\mu$m corresponding to the three considered types of structures denoted by means of $\rm{S_{1}}$, $\rm{S_{2}}$ and $\rm{S_{3}}$ abbreviations. \label{initial_parameters}}
\end{table}

\section{Results}

In order to assess the possibility of changing the Poisson's ratio of the considered graded mechanical metamaterial, three different quasi-2D structures presented in Fig. \ref{model_fig}(b) are analyzed. The main difference between them is the number of structural elements corresponding to substructures characterized by the negative and positive Poisson's ratio respectively. Thus, it is expected that upon changing the ratio of structural units associated with these two substructures, it may be possible to adjust the resultant Poisson's ratio of the entire structure.

\subsection{Quasi-2D mechanical metamaterials}

\begin{figure*}
	\centering
	\includegraphics[width=0.99\textwidth]{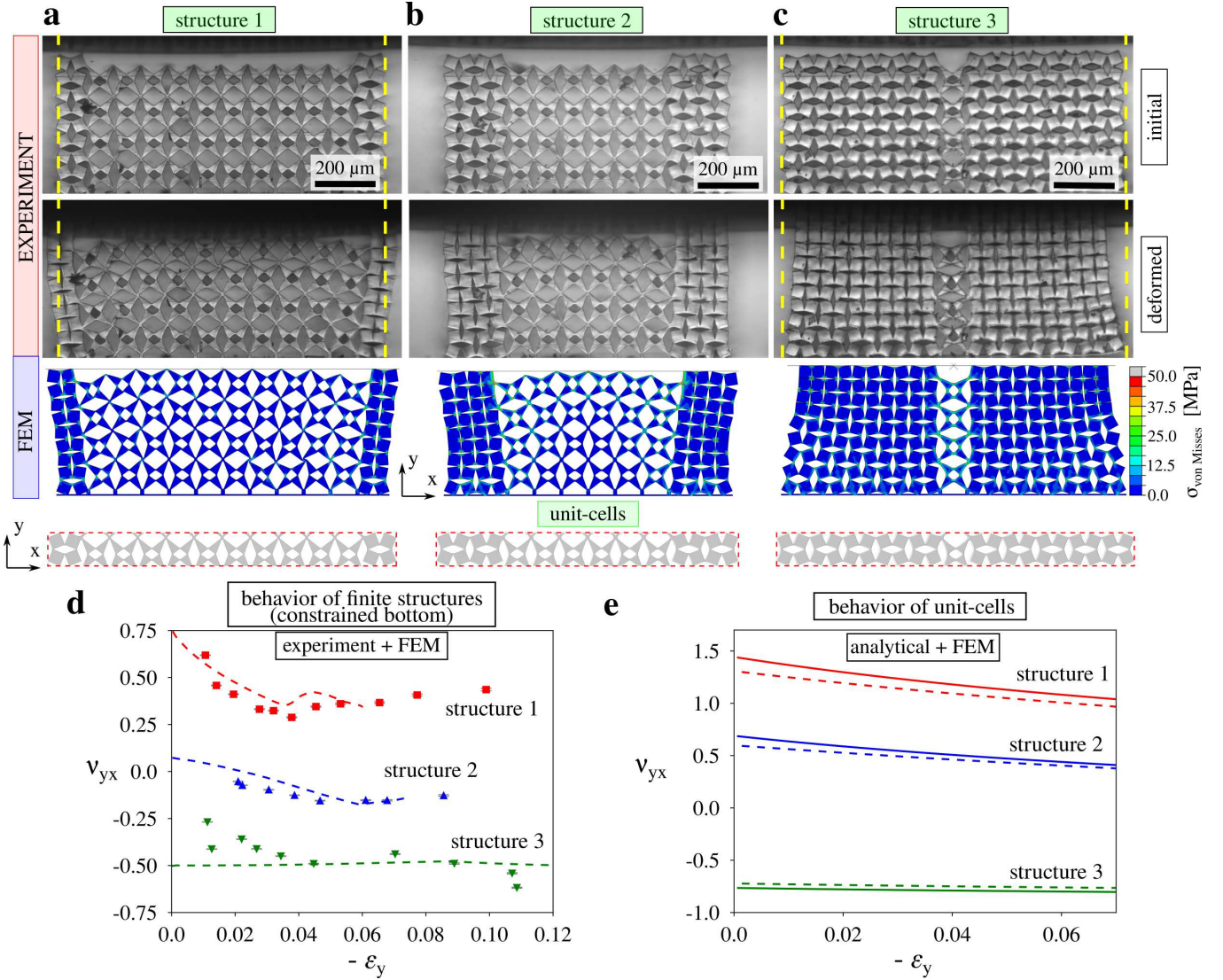}
	\caption{Behavior of finite quasi-2D structures subjected to the compression along the $y$-axis. {\bf a)} Mechanical deformation of Structure 1 corresponding to the $N_{\rm{S,x}}$=1 and $N_{\rm{T,x}}$=10 parameters. {\bf b)} Behavior of Structure 2 associated with $N_{\rm{S,x}}$=2 and $N_{\rm{T,x}}$=6 parameters. {\bf c)} Compression of Structure 3 corresponding to $N_{\rm{S,x}}$=5 and $N_{\rm{T,x}}$=1. {\bf d)} Engineering Poisson's ratio corresponding to the compression of the considered finite structures (see Fig. \ref{model_fig}(b)) in the $xy$ plane along the $y$ axis. {\bf e)} Engineering Poisson's ratio corresponding to individual unit-cells with periodic boundary conditions determined through the use of the analytical model and FEM simulations. In this case, the dashed and solid lines correspond to the FEM simulations and the analytical model respectively.}
	\label{results_2D}
\end{figure*}

As shown in Fig. \ref{results_2D}(a), in the case of the first of the analyzed structures, i.e. Structure 1, the top part of the system which is unconstrained expands significantly with respect to its initial dimension upon being compressed. This in turn is an indication of the positive Poisson's ratio. However, as mentioned in the Experiment subsection, in order for the results not to be strongly influenced by the edge effects at the top of the compressed system, the second topmost unit-cell is taken into consideration to assess mechanical properties of the structure (see highlighted unit-cell in Fig. \ref{model_fig}(a)). As can be seen based on the provided pictures and images extracted from FEM simulations, the horizontal dimension of this specific unit-cell still expands significantly with respect to its initial length. This, in addition to the shrinkage of the unit-cell along the $y$-axis, results in the positive Poisson's ratio behavior. Furthermore, it should be noted that the specific way how the Poisson's ratio was calculated for finite systems analyzed in this work is provided in the Supplementary Information.

According to Fig. \ref{results_2D}(b), the behavior of the system is significantly changed in comparison to what was observed for Structure 1. More specifically, it appears that the change in the horizontal dimension of the structure is very small. This is indicative of the resultant Poisson's ratio being in the vicinity of zero.

As presented in Fig. \ref{results_2D}(c), the last of the analyzed finite systems, i.e. Structure 3, exhibits a drastically different mechanical response than structures 1 and 2. More specifically, upon having a closer look at the generated results, it is possible to note that the horizontal dimension (measured along the $x$-axis) of the analyzed part of the system was significantly decreased as a result of the compression along the $y$-axis. This in turn is indicative of the highly negative Poisson's ratio exhibited by the structure. Nevertheless, to gain a better understanding of the effect that the change in a number of structural elements of individual substructures has on the overall behavior of the resultant graded system, it is necessary to analyze quantitative results.

According to Fig. \ref{results_2D}(d), it is clearly visible that depending on the number of structural elements corresponding to the two types of substructures constituting the system, it is possible to very significantly change the resultant Poisson's ratio of the entire system. In fact, it is even possible to change it from highly positive to highly negative values which is certainly very appealing from the perspective of the future design of complex multi-substructure graded mechanical metamaterials. Furthermore, it should be emphasized that all observations made during the above qualitative analysis are confirmed by the quantitative results. In addition, both experimental results and results extracted from FEM simulations appear to be in a very good agreement which confirms their validity. Upon having a closer look at the generated results, it is possible to gain a better understanding of how the individual structures behave during the compression process. In the case of results associated with Structure 1, it is possible to observe that the Poisson's ratio assumes positive values throughout the entire deformation process. More specifically, this quantity initially assumes values around 0.75. However, upon exceeding the strain of magnitude 0.02, the Poisson's ratio seems to maintain the value from the interval between 0.3 and 0.5. Furthermore, in the case of Structure 2, it is possible to note that the Poisson's ratio initially assumes a small positive value in order to later exhibit the Poisson's ratio around -0.1 upon reaching larger strains. This is a great indication that in general, one can design the considered graded mechanical metamaterial in a way where the resultant Poisson's ratio will remain in the vicinity of zero which can be important in the case of applications utilizing mechanical materials that are supposed to retain their lateral dimension upon being deformed. In the case of the last of the analyzed systems, i.e. Structure 3, one can note that the value of the Poisson's ratio is maintained in the vicinity of -0.5 throughout the entire deformation process. This, in turn, is a very interesting result from the point of view of devices that utilize materials that are not supposed to lose their auxeticity throughout the deformation (e.g. protective equipment).

Even though the experimental and computational results presented in Fig. \ref{model_fig}(d) are very promising, it is essential to check how the considered systems would behave if one was to form a very large structure composed of unit-cells proposed in this work instead of using relatively small finite structures. To this aim, the behavior of the analytical model corresponding to the highly idealized pin-jointed hinges was compared to the FEM simulations corresponding to the single unit-cell with realistic hinges (see Fig. \ref{model_fig}(a)) and periodic boundary conditions being imposed. As shown in Fig. \ref{results_2D}(e), similarly to the results generated for finite structures, it is possible to observe a very large variation in the resultant Poisson's ratio depending on the type of the structure. However, it is also possible to note that in this case, Poisson's ratios associated with Structure 1 and Structure 2 assume more positive values than was formerly the case. Similarly, the Poisson's ratio corresponding to Structure 3 assumes more negative values throughout the deformation process than was observed for the formerly analyzed finite system. Finally, it is worth noting that both analytical results associated with idealized structures and FEM simulations lead to almost identical results. This in turn is an indication that relatively small changes in the thicknesses of hinges used in this work are not expected to influence the results. In addition, one can note that all of the results presented in the main text correspond to the compression of the analyzed samples. This stems from the fact that it is extremely difficult to stretch micro-scale structures. However, one can conduct such an analysis in the case of the FEM simulations. As shown in Fig. S3 in the Supplementary Information, trends corresponding to Poisson's ratio exhibited by the considered systems are unaffected by the change in the loading direction. This originates from the fact that all of the considered samples are unimode structures, i.e. their deformation can be described through the change in only one parameter.

\subsection{3D structures}

The concept reported in this work is not limited to 2D or quasi-2D systems and in general can be extended to 3D structures capable of exhibiting the desired mechanical properties in multiple directions. In order to show it, in this work, we designed two new structures that are based on quasi-2D systems.

\begin{figure*}
	\centering
	\includegraphics[width=0.99\textwidth]{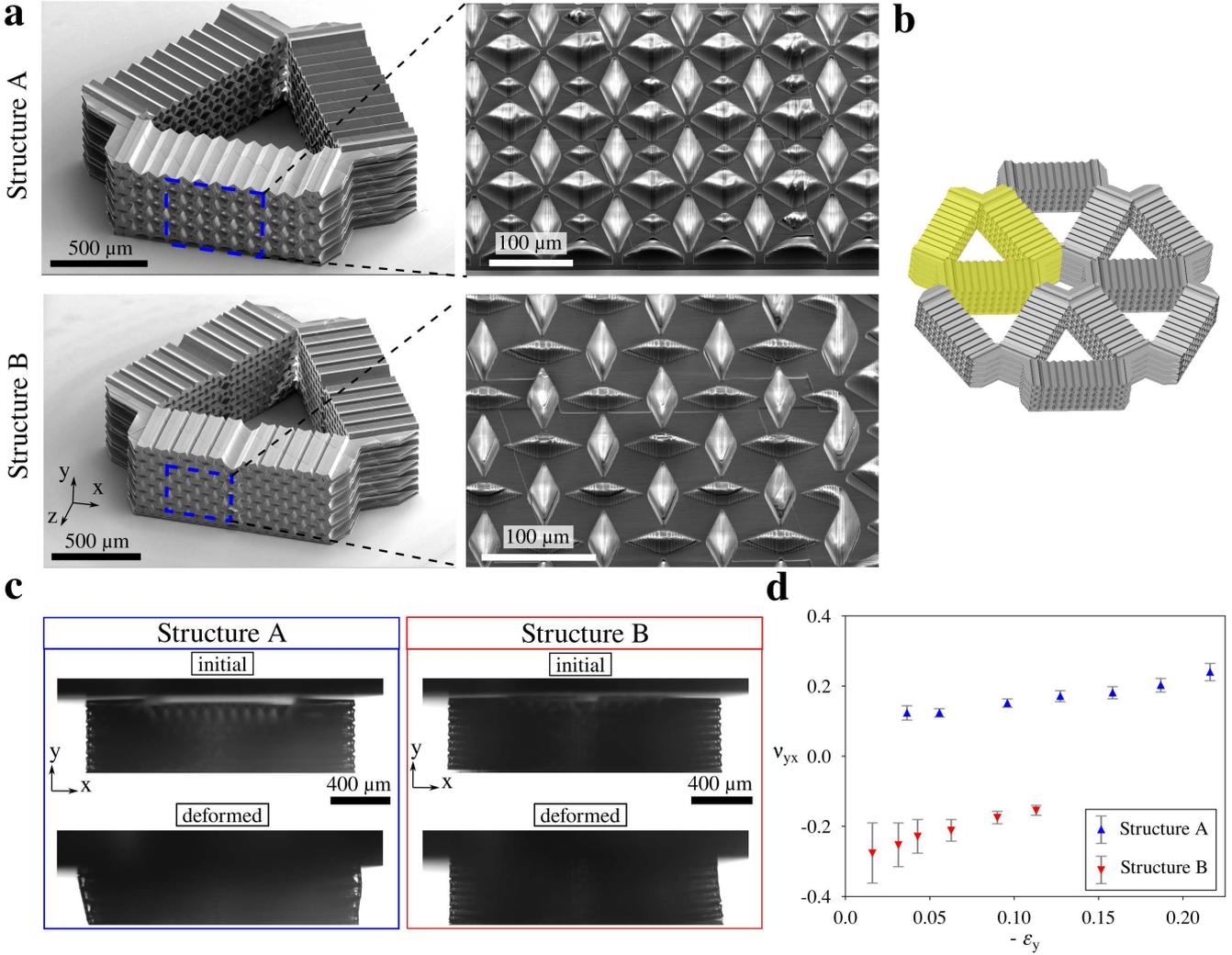}
	\caption{The behavior of finite 3D structures subjected to the compression along the $y$-axis. {\bf a)} SEM images of the two types of 3D structures analyzed in this work. {\bf b)} A conceptual diagram showing how the considered 3D structures could become a part of a larger hexagonal lattice. {\bf c)} Pictures taken by means of the optical camera showing the change in the cross-section of the analyzed structures in the $xy$ plane upon being compressed along the $y$-axis. {\bf d)} The graph showing the engineering Poisson's ratio corresponding to the analyzed 3D structures. In this case, the horizontal dimension used in the calculations of the Poisson's ratio corresponds to the topmost part of the finite system.  }
	\label{results_3D}
\end{figure*}

The considered 3D structures are called Structure A and Structure B and are presented in Fig. \ref{results_3D}(a). These mechanical metamaterials are based on quasi-2D structures named Structure 1 and Structure 3 respectively that were already proven to result in very different Poisson's ratios. Upon taking Structure A as an example, it should be emphasized that in order to construct it, three indentical elements in the form of the quasi-2D Structure 1 were used that were then spatially arranged to form a configuration resembling an equilateral triangle (in terms of its cross-section) in the $xz$ plane. As shown in Fig. \ref{results_3D}(a), the corners of such three blocks were also connected together by means of additional layers of a material to ensure that they move together during the deformation process of the entire system. Similarly, Structure B was designed based on three block-like elements corresponding to Structure 3.

As shown in Fig. \ref{results_3D}(b-c), as Structure A is being compressed, its horizontal dimension increases which results in the positive Poisson's ratio. It should be noted that in this case, the horizontal and vertical dimensions used in the calculation of the Poisson's ratio correspond to the horizontal dimension of the topmost part of the cross-section shown in Fig. \ref{results_3D}(c) and the overall vertical dimension respectively. Conversely, in the case of Structure B, the system exhibits auxetic behavior throughout the deformation process. Both of these results should be expected based on properties of quasi-2D samples referred to as Structure 1 and Structure 3. However, one can note that for 3D structures, the magnitude of the Poisson's ratio is significantly smaller than was formerly the case. This stems from the fact that the horizontal dimension of the cross-section of the system incorporates additional non-deformable elements that were used to connect the adjacent structural blocks together. At this point, it is also worth noting that even though this is not the subject of investigation in this work, the regions of maximum stress accumulation within the structures can be ascertained based on the already conducted FEM simulations of the behavior of 2D structures. More specifically, the maximum stress can be observed at the hinges connecting mutually-rotating structural elements. On the other hand, the localized stresses within the central regions of the square-shaped or triangle-shaped structural elements are negligible.

\subsection{Wave propagation}

In addition to mechanical properties, one can also analyze the effect that the variation in the composition of considered mechanical metamaterials has on the wave propagation through these structures. The phonon dispersion relation was calculated using COMSOL Multiphysics software for the three types of unit-cells presented in Fig. \ref{results_2D}. To generate the results, in the case of all of the analyzed examples, Floquet periodic boundary conditions were implemented in the $x$- and $y$-directions assuming two-dimensional unit-cells. 

\begin{figure*}
	\centering
	\includegraphics[width=0.99\textwidth]{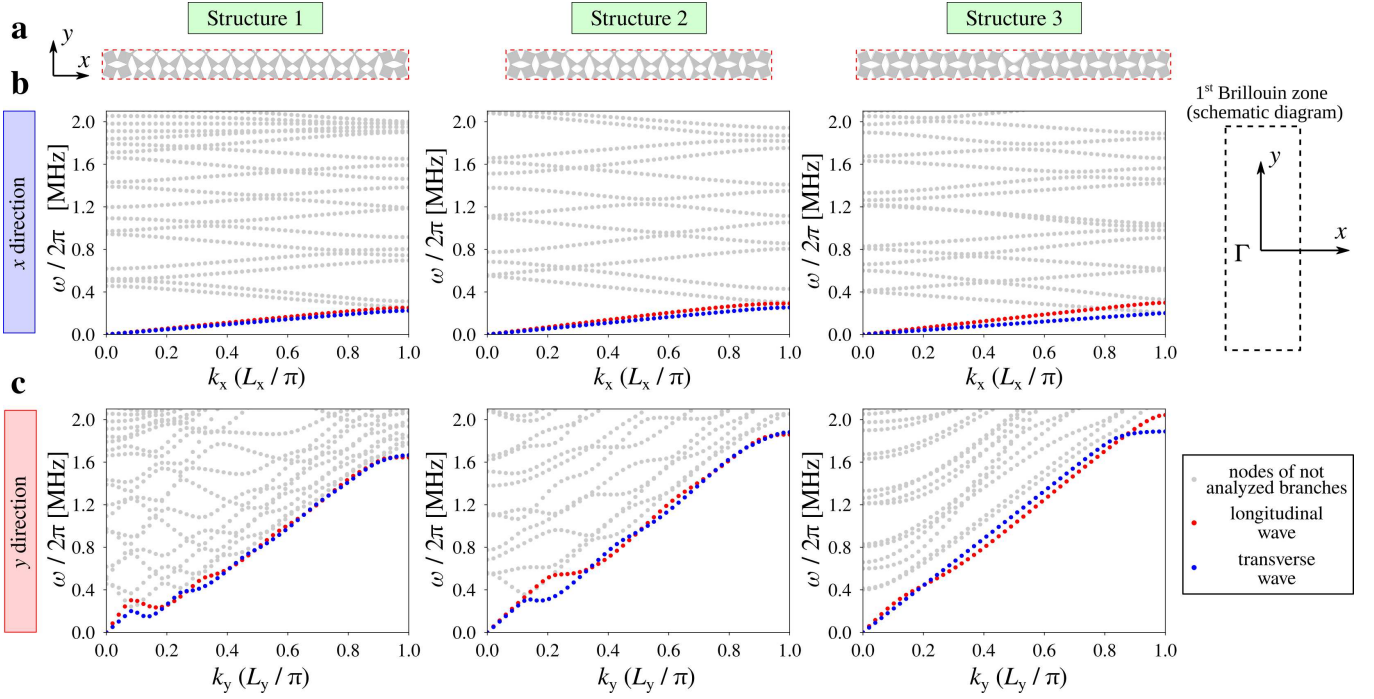}
	\caption{Phonon dispersion derived for unit-cells corresponding to the considered structures. {\bf a)} Unit-cells of considered structures (in real space). {\bf b)} Phonon dispersion graphs corresponding to the $x$ direction. {\bf c)} Phonon dispersion graphs associated with the $y$ directions. For all graphs, there is always one band highlighted by means of the red and blue colors. These bands were used in the quantitative analysis to determine the group velocities for each of the structures. The auxiliary diagram of the first Brillouin zone is not up to scale.}
	\label{wave_results}
\end{figure*}

The main objective of this part of the work is to determine the group velocities associated with longitudinal and transverse waves propagating in two principal directions, i.e. along the $x$ and $y$ axes. This will give an idea of how does the speed of the wave propagation change upon considering systems composed of a different number of auxetic and positive Poisson's ratio structural elements. This, in turn, can prove to be interesting for example in the case of sensors capable of detecting the configuration assumed by the system in the case when it is not easy to determine it via a direct optical observation. Such possibility is of particular significance for micro-scale structures where such procedure could prove to be challenging. In addition, it should be mentioned that for each of the structures, the results were generated for the 20 first eigenmodes. However, according to the generated data, no band gaps were observed. Instead, as shown in Fig. \ref{wave_results}, one can note a significant hybridization originating from the complexity of the resultant structures and their general design based on 2 effectively different metamaterials. Finally, one can note that the observed results correspond to the Megahertz range which stems from the fact that the analyzed structures were constructed at the micro-scale. However, in general, together with the size of the structure itself, the frequency range can be also adjusted to match the desired wavelength. At this point, one should also note that in our analysis, we are only focusing on the behavior of the 2D structures. This stems from the fact that primitive cells associated with the 3D structures portrayed in Fig. 3 have to be defined in 3D and are very complex. Thus, the corresponding numerical simulations would be very time-consuming and difficult. At the same time, we do not expect for the conclusions related to the wave propagation in 3D structures to be significantly different from those associated with 2D systems. In addition, it should be noted that the presented results are valid for specific structures investigated in this work. However, should someone fabricate similar samples at a different scale or through the use of a different material, the band structure associated with the system would be expected to be rescaled. Thus, to obtain more general results, one can normalize the presented frequencies with respect to the lattice constant and the P wave velocity. These results are provided in the Supplementary Information in a form of Fig. S2 and correspond to the same trends as shown in Fig. \ref{wave_results}.

In order to determine group velocities, only the first two branches were taken into consideration for each of the graphs presented in Fig. \ref{wave_results}. These branches are graphically highlighted by means of the blue and red colors that correspond to transverse and longitudinal waves respectively that propagate in a given direction. In addition, since the group velocity is in general defined as $\upsilon_{g} = \frac{d\omega}{d k}$ ($\omega$ - angular frequency, $k$ - wave vector), only the initially linear fragments of these branches were considered in order to calculate respective group velocities. At this point, it should also be noted that overall, phonon dispersion graphs for the wave propagation in the $x$ and $y$ directions is very different. In fact, this stems from the high anisotropy of the considered structures.

\begin{table}
	\renewcommand{\arraystretch}{1.35}
	\begin{tabular}{ |c|c|c|c|c|}	
		
		\hline
		\multicolumn{2}{|c|}{} & S1 & S2 & S3 \\  \hline
		\multirow{2}{*}{$x$ direction} & $\upsilon_{g}^{L}$ [m/s] & 596.4 & 620.0 & 702.9\\ 
		\cline{2-5} & $\upsilon_{g}^{T}$ [m/s]& 520.6 & 514.6  & 460.2 \\  \hline
		\multirow{2}{*}{$y$ direction} & $\upsilon_{g}^{L}$ [m/s]& 781.9 & 567.9 & 574.3\\ 
		\cline{2-5} & $\upsilon_{g}^{T}$ [m/s]& 515.0 & 520.6 & 461.8 \\  \hline
		
	\end{tabular}
	\caption{Values of group velocities $\upsilon_{g}^{L}$ and $\upsilon_{g}^{T}$ corresponding to longitudinal and transverse waves respectively. Abbreviations S1, S2 and S3 stand for Structure 1, Structure 2 and Structure 3. \label{group_velocities}}
\end{table}

As shown in Table \ref{group_velocities}, in the case of waves propagating in the $x$-direction, the group velocity associated with longitudinal waves ($\upsilon_{g}^{L}$) seems to be the smallest for Structure 1, in order to reach the highest value of 702.9 m/s in the case of Structure 3. This means that in this specific direction, it is possible to observe a significant change in $\upsilon_{g}^{L}$ upon changing the composition of the structure. Conversely, in the case of transverse waves propagating in the $x$-direction, the group velocity $\upsilon_{g}^{T}$ gradually decreases with a relatively small change in the magnitude. Furthermore, it should be emphasized that for waves propagating in the $y$-direction, it is hard to observe any specific trends related to the change in the magnitude of the group velocities corresponding either to longitudinal or transverse waves. In addition, as shown in Fig. \ref{wave_results}(b), the branches form a rather perturbed picture in comparison to what could have been observed for waves propagating in the $x$-direction. This can be explained by the fact that in the $y$-direction, the substructures constituting the considered systems form parallel stripes corresponding to a very different topology. Hence, it is possible to expect a strong interference that influences the results. Furthermore, it should be noted that trends presented in Fig. \ref{wave_results} and Table \ref{group_velocities} can also be observed if one was to replace the two types of substructures considered in this work with bulk materials having the same external dimensions and effective mechanical properties (see Supplementary Information). This indicates that the presented results are reliable and that similar types of the complex phonon dispersion relations could potentially be observed for other multi-substructure graded structures. Last but not least, we note that both longitudinal and transverse bands for Structure 1 show some very important non-monotonic behavior where its form was recently reported as the roton effect in metamaterials. The origin of this effect was attributed to the third-order interactions in equivalent mass-springs models \cite{chen2021roton, iglesias2021experimental}. Finally, is should be noted that further details related to the analysis of the observed results are provided in the Supplementary Information. Amongst others, the phonon dispersion relation corresponding solely to geometries related to rotating squares and rotating triangles mechanical metamaterials are presented. This, in turn provides, an information about the wave propagation properties of materials constituting individual part of the more complex graded structure considered in this work.

\section{Discussion }

The novel graded mechanical metamaterial proposed in this work offers numerous advantages over a majority of known mechanical metamaterials and other graded structures. The considered graded system incorporates two structures that can have their deformation fully described by a change in one parameter assuming relative rigidity of their structural elements. As a result, the evolution of the entire system investigated in this work can be also described by a variation in one geometric parameter which distinguishes it from many other graded structures \cite{Mirzaali_Zadpoor_2020, Guo_Huang_2021, Han_Kei_2022, NIKNAM2020109129}. Such unique design, also results in the very high stability of the system which is often not observed for other graded metamaterials composed for example of different re-entrant structural motifs. In addition, the two types of substructures present within the system are characterized by drastically different Poisson's ratio. Hence, depending on the composition of the considered mechanical metamaterial, it can exhibit either very positive or very negative values of this quantity. This, in turn, as demonstrated in another study \cite{Dudek_2022_Adv_Mater}, can enable the structure to exhibit a complex shape-morphing behavior. Finally, it should be mentioned that in terms of the manufacturing process, the proposed design makes it possible to 3D print different complex components of the structure at the micro-scale with a very high precision. However, In the case of many other designs known in the literature, this task would become significantly more challenging due to the presence of overhanging elements.

All of the results presented in this work suggest that the considered mechanical metamaterials can exhibit a very broad range of the Poisson's ratio including both highly negative and positive values depending solely on the ratio of the number of structural elements corresponding to the negative and positive Poisson's ratio part of the system. It means that within certain limits, one can achieve an arbitrary value of this parameter based solely on two types of substructures. This, in turn, is of great significance as it may enable engineers to produce relatively simple mechanical materials that can exhibit desired mechanical properties without the need of utilizing very complex and unique designs for each type of application. It is also worth mentioning that the latter approach is relatively expensive and may lead to malfunctions due to the unnecessary complexity of structural designs. In addition, in this study, it was demonstrated that the considered results can be observed in the case of microscopic systems. This means that in addition to applications typical for many auxetic mechanical metamaterials, e.g. protective devices or functional materials utilizing an enhanced fracture toughness, indentation resistance as well as strain-dependent fully-reversible porosity and deformation,  this concept may also be implemented in the case of novel biomedical devices and flexible electronics. Particularly important in this regard is the potential ability of structures presented in this work to exhibit complex shape morphing upon connecting multiple elements similar to considered structures in order to form larger lattices. As a result, one can obtain a graded metamaterial capable of exhibiting very different local values of the Poisson's ratio which, as presented in the recent studies devoted to micro-scale shape-morphing mechanical metamaterials \cite{Dudek_2022_Adv_Mater}, may lead to the reconfiguration of the initial system into a complex predefined shape. Such ability could for example prove to be useful in the design of efficient stents where upon being subjected to the external stimulus, devices based on the proposed concept could provide a better support to specific parts of the blood vessel.

\section{Conclusion}
In this work, novel 2D and 3D functionally graded mechanical metamaterials were proposed that consist of two substructures corresponding to the negative and positive Poisson's ratio. It was shown that depending on the number of structural elements corresponding to each type of the substructure, it is possible to significantly change the resultant value of Poisson's ratio to reach its desired magnitude. It was also shown that structures investigated in this work can be constructed at the microscale in addition to larger scales that are typical for mechanical metamaterials. This means that the proposed concept could be used for example in the design of the biomedical devices such as stents where the proposed solutions could enhance the efficiency of the already existing devices by locally supporting specific parts of the blood vessel at scales that normally cannot be achieved in the case of mechanical metamaterials. In addition, results reported in this study could also be utilized at a larger scale in the design of efficient protective devices or sports equipment corresponding to a specific value of the Poisson's ratio that is beneficial from the perspective of the specific application. Finally, it was demonstrated that the change in the composition of considered structures affects the wave propagation through the system. This in turn may be interesting from the point of view of artificial materials capable of exhibiting a local wave attenuation.




%
%




\section*{Declaration of competing interest}
The authors declare that they have no known competing financial interests or personal relationships that could have appeared to influence the work reported in this paper.

\section*{Acknowledgements}
\indent
This work was partly supported by the french RENATECH network and its FEMTO-ST technological facility.\\
\indent
K.K.D. acknowledges the support of the Polish National Science Centre (NCN) in the form of the grant
awarded as a part of the SONATINA 5 program, project No. 2021/40/C/ST5/00007 under the name “Programmable magneto-mechanical metamaterials guided by the magnetic field”.\\
\indent
This research was funded by the Polish Minister of Education and Science under the program “Regional Initiative of Excellence” in 2019–2023, project
No. 003/RID/2018/19, funding amount PLN 11 936 596.10. aa

\section*{Data Availability}
The data presented in this work are available from the corresponding authors uopn reasonable request.

\bibliographystyle{unsrt}

\bibliography{manuscript_references.bib}

%

\end{document}